\documentclass[aps,prl,twocolumn,groupedaddress,floatfix]{revtex4}

\usepackage{graphicx}
\usepackage{amssymb}

\begin{document}

\title{Lattice Dynamics and Melting of a Nonequilibrium Pattern}
\author{Daniel I. Goldman}
\email[]{goldman@chaos.utexas.edu}
\homepage{http://chaos.ph.utexas.edu/~goldman}
\author{M. D. Shattuck}
\author{Sung Joon Moon}
\author{J. B. Swift}
\author{Harry L. Swinney}
\affiliation{Center for Nonlinear Dynamics,
                The University of Texas at Austin,
                Austin, TX 78712}
\date{\today}

\begin{abstract} 

We present a new description of nonequilibrium square patterns as a harmonically coupled crystal lattice. In a vertically oscillating granular layer, different transverse normal modes of the granular square-lattice pattern are observed for different driving frequencies ($f_d$) and accelerations. The amplitude of a mode can be further excited by either frequency modulation of $f_d$ or reduction of friction between the grains and the plate. When the mode amplitude becomes large, the lattice melts (disorders), in accord with the Lindemann criterion for melting in two-dimensions.

\end{abstract}

\pacs{}

\maketitle

Systems driven away from thermodynamic equilibrium often form patterns when forced beyond a critical threshold. Close to this bifurcation, the dynamics of the nonequilibrium patterns are well described by partial differential equations called amplitude equations, whose forms are universal~\cite{croAhoh}. However, these equations lose predictive power for larger forcing. As an alternative description of patterns, Umbanhowar {\em et al.}~\cite{umbAmel96} conjectured that interacting localized structures, ``atoms" of the patterns, could be the building blocks of spatially extended patterns. In this Letter, we demonstrate that this speculation was correct: we describe and {\em predict} the behavior of a pattern in a particular system by treating it as a finite number of interacting elements. Such an approach can replace the description of patterns by partial differential equations with a possibly simpler description based on a finite set of coupled ordinary differential equations. 

We use a well-known pattern forming system, a thin layer of oscillated granular material. We show that the dynamics (including all oscillatory motions and disordered states) of square patterns are analogous to the dynamics of a discrete lattice of harmonically coupled elements. The descriptive and predictive power of such a framework is especially useful for the patterns formed in granular materials because the governing equations are only now being rigorously tested~\cite{grantest}, and the exact forms of the amplitude equations are not yet established.

\begin{figure}[h!tb]
\includegraphics[width=3.25in]{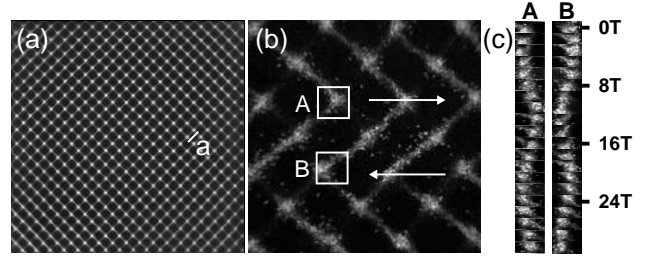}
\caption{\label{largelattice}
(a) A lattice pattern strobed at $f_{d}/2$ and time averaged ($\Gamma=2.90$ and $f_{d}=25$ Hz, parameters for which the lattice is almost motionless). The image shows the entire container, $18 \times 18~\mbox{cm}^2$. (b) Close up snapshot image ($3 \times 3~\mbox{cm}^2$) at $\Gamma=2.90$ and $f_{d}=30$ Hz, for which the lattice oscillates vigorously. (c) The evolution in time (with units $T=1/f_d$) of the peaks in boxes A and B in (b); these peaks oscillate out of phase with a frequency about 10 times smaller than $f_{d}/2$.}
\end{figure}

{\em Experiment.} A layer of $0.17$~mm diameter bronze spheres was oscillated vertically in an evacuated cell at driving frequency $f_{d}$ with non-dimensional peak plate acceleration $\Gamma=A(2\pi f_d)^2/g$, where $A$ is the amplitude of the plate oscillation and $g$ is the gravitational acceleration. For the layer depth studied (4 particle diameters), square patterns formed for $2.5 < \Gamma < 4.0$ and $f_{d} < 36$ Hz.  The granular surface was imaged using low angle illumination that created bright regions at the peaks~\cite{melAumb}. The scattered light was collected by a $256 \times 256$ pixel CCD camera. 

{\em Lattice pattern.} The square patterns, illustrated in Fig.~\ref{largelattice}(a), oscillate subharmonically at $f_{d}/2$; after each plate oscillation, a peak becomes a crater. At the phase in the plate oscillation cycle when the pattern amplitude is maximum, the pattern is composed of an array of peaks arranged in a square lattice connected by a network of thin lines of particles. At maximum amplitude each peak typically contains several hundred particles. Images are collected at this phase in the cycle. In the dark regions between the peaks, there are almost no grains. Thus, when strobed at $f_{d}/2$, the pattern resembles a two-dimensional (2D) square crystal lattice made of discrete elements separated by lattice constant $a$. In this paper, we only consider the motion of the lattice pattern strobed at $f_{d}/2$.

{\em Lattice oscillation.} We will now show that the lattice analogy is more than superficial: the square patterns also exhibit the {\em dynamics} of a coupled lattice. In general, the center of mass of each peak (lattice element) oscillates around its equilibrium (time-averaged) lattice site. We find that for a range of control parameters, the oscillation is periodic; a particular example is illustrated in Fig.~\ref{largelattice}(b) and (c). The peaks in a row at an angle $\pi/4$ to the natural lattice direction maintain a constant separation of $\sqrt{2}a$ as they oscillate. Such an oscillation of the pattern is analogous to a single excited transverse normal mode in the (1,1) direction of a 2D crystal lattice. We find that all periodic motions of the lattice resemble $(1,1)_T$ modes (or the degenerate $(1,-1)_T$), with different wavevectors and oscillation frequencies. 

\begin{figure}[h!tb]
\includegraphics[width=3in]{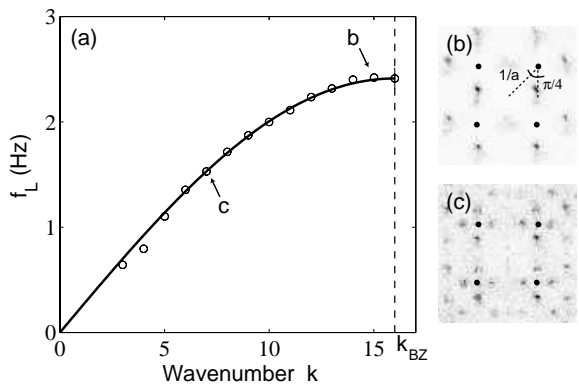}
\caption{\label{dispersion}
(a) Comparison of the measured $(1,1)_T$ dispersion relation ($\circ$) at $f_{d}=25$ Hz and $\Gamma=2.75$ with a lattice model (solid line). The wavenumber $k$ is in units of $4\pi/({\sqrt{2} a N})$, where $a$ is the lattice constant and $N=34$ is the number of $(1,1)$ rows in our square container ($N=32$ in Fig.~\ref{largelattice}). Only $N/2$ positive $k$ modes are shown since the lattice oscillations are decomposed into periodic Fourier components. The dashed line denotes the edge of the first Brillouin zone ($k=k_{~\mbox{BZ}}$) for the $(1,1)_T$ modes.  (b) and (c) show the modulus of the spatial Fourier transforms, $|\tilde{I}(k_x,k_y,f_{L})|$, at two lattice oscillation frequencies, $f_{L}=2.3$ Hz and $f_{L}=1.2$ Hz, respectively. The sidebands represent the spatial modulation of the lattice $\pi/4$ from the basic square lattice direction---the $(1,1)_T$ modes. The four peaks ($\bullet$) that form the basic square lattice (found at $f_{L}=0$ Hz) have a power about 50 times larger than the background.}
\end{figure}

{\em Dispersion relation.} The evidence that the granular lattice behaves like a discrete coupled lattice is provided by the dispersion relation shown in Fig.~\ref{dispersion}(a). Since the lattice dynamics is in general more complicated than the periodic oscillation of a single mode shown in Fig.~\ref{largelattice}, the dynamical behavior is determined from the three-dimensional discrete Fourier transform of a time series of images, giving $~\tilde{I}(k_{x}, k_{y}, f_{L})$. Only certain values of oscillation frequencies $f_L$ and wavevectors $(k_x, k_y)$ contain power ($|\tilde{I}|^2$), and these combinations yield the dispersion relation for the lattice in the $(1,1)_T$ (or the degenerate $(1,-1)_T$) direction~\cite{modenote}.

As a first approximation to the unknown 2D potential between the lattice elements, following Kittel~\cite{kittel}, we assume that the modes can be decoupled in different crystal directions. Since we observe only the $(1,\pm1)_T$ modes, we compare the data to a 1D lattice model in which we assume that the transverse motion of each $(1,1)$ row is harmonically coupled to its nearest $(1,1)$ neighbors. As shown in Fig.~\ref{dispersion}(a), the measured dispersion relation is well described by the dispersion relation of the 1D lattice model with $f_{L}=f_{BZ}|\sin(ka/(2\sqrt{2})|$, where $f_{BZ}$ is the frequency at the edge of the Brillouin zone (proportional to the square root of the ratio of effective spring constant to the effective mass of a $(1,1)$ row), $a$ the lattice constant, and $k$ is the magnitude of a wavevector in the $(1,1)$ direction~\cite{louck}. We have found that for a wide range of $\Gamma$ and $f_d$, the measured dispersion relations are fit well by the harmonically coupled lattice model; as $\Gamma$ and $f_{d}$ are changed, $f_{BZ}$ varies between $1.5$ Hz and $2.5$ Hz, an order of magnitude smaller than $f_{d}$.

\begin{figure}[h!tb]
\includegraphics[width=3in]{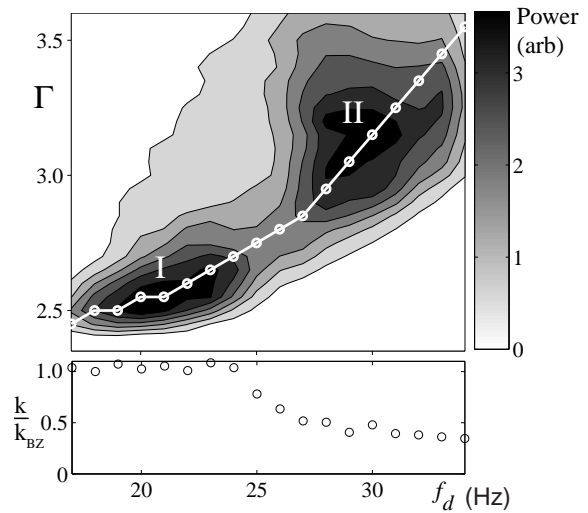}
\caption{\label{resplane}
Normal modes of the lattice exhibit two regions of intense excitation, I and II. Top panel: the grayscale indicates the power in the dominant mode, which was obtained by integrating the spectral power above a noise background. Bottom panel: The wavevector of the dominant mode. The points are taken along the path in the top panel.}
\end{figure}

{\em Excitation of normal modes.} We have demonstrated that lattice dynamics describes the square patterns; therefore, we can view the periodic oscillations of the pattern as individually excited normal modes. In two ranges of $\Gamma$ and $f_d$, the oscillation of the pattern becomes very intense (Fig.~\ref{resplane}). The two regions are characterized by excitation of specific normal modes: In region I, the power is dominant in the mode at the edge of the $(1,1)$ Brillouin zone, $k=k_{BZ}=2\pi/(\sqrt{2}a)$, while in region II, the power is dominant in a mode near the middle of the Brillouin zone. We do not know why these particular modes are excited. Away from the two regions of excitation, the lattice is nearly stationary, with small amplitude oscillations of the lattice elements around the equilibrium sites. In these quiescent regions, the total power is small and is roughly independent of the mode number: the lattice acts as if it is in contact with a thermal bath.

\begin{figure}
\includegraphics[width=3in]{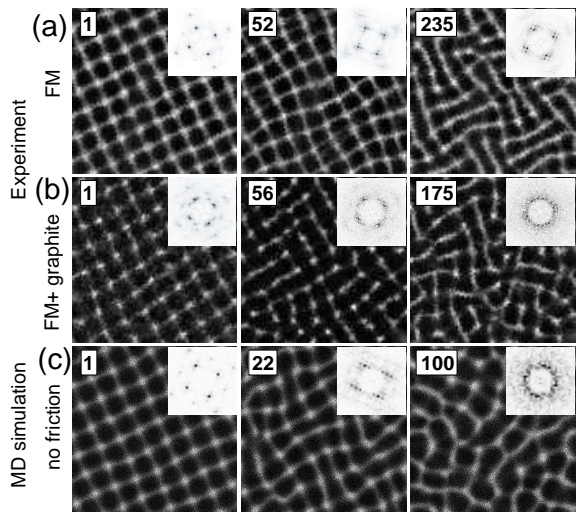}
\caption{\label{comparison} Defect creation and melting for three cases after a sudden change in the system parameters.  The numbers are the number of oscillations after the change in conditions, and the insets show the structure factor (power spectrum) corresponding to the pattern. (a) At $t=0$, frequency modulation with $f_{mr}=2$ Hz and $f_{ms}=5$ Hz was applied ($\Gamma=2.9$, $f_d=32$ Hz). (b) At $t=0$, the same frequency modulation as in (a) was applied for particles which had been cleaned and to which graphite had been added. (c) Molecular dynamics simulation: at $t=0$, the friction coefficient $\mu$ between the grains and the plate was reduced from $0.5$ to $0$ ($\Gamma=3.0$ and $f_d=32$ Hz, for which the pattern weakly oscillates when $\mu=0.5$).}
\end{figure}

{\em Defect formation.} In a real crystal lattice, if the amplitude of the oscillations is large enough, defects form and melting occurs. We see the analogous effect in the square pattern lattice: near region II, the single mode oscillations are large enough to locally break the lattice and form defects, which either remain in the lattice or travel through the crystal and annihilate at the boundary. Thus, in region II the patterns do not display perfect long range order.

The defect creation rate can be enhanced by resonantly exciting a specific normal mode, as shown in Fig.~\ref{comparison}(a). We achieve this by a slow frequency modulation of the layer oscillation: the signal applied to the container has the form $y=A\sin(2\pi f_{d}t+\frac{f_{ms}}{f_{mr}}\sin{2\pi f_{mr}t})$, where $f_d$ is modulated at a rate $f_{mr}$ with a modulation span $f_{ms}$. The lattice responds to the frequency modulation by oscillating in a $(1,1)_T$ normal mode with a frequency of exactly $f_{mr}/2$, and the strength of the response increases as $f_{ms}$ increases. Several hundred oscillations after the modulation is turned on, the amplitude of the excited mode becomes quite large, and a few defects are created as the mode locally shears the lattice apart. 

\begin{figure}[h!tp]
\includegraphics[width=3in]{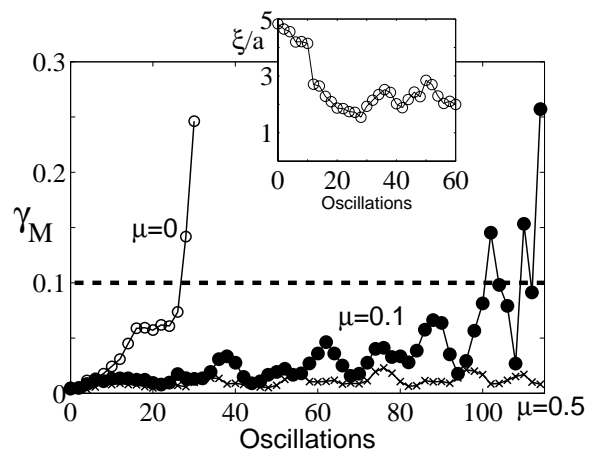}
\caption{\label{decorr}
Melting occurs when the Lindemann ratio $\gamma_M$ (see text) reaches approximately $0.1$. Main figure: the evolution of $\gamma_M$ in simulations with $\Gamma=3.0$ and $f_d=32$ Hz after the friction coefficient $\mu$ was changed at $t=0$ from $\mu=0.5$. The inset shows that the correlation length of the pattern for $\mu=0$ reached the plateau region when $\gamma_M \approx 0.1$.}
\end{figure}

{\em Friction and melting.} To further increase the rate of defect formation, we have reduced the friction between the particles and between the particles and the plate by adding approximately $0.2~\mbox{grams}$ of graphite powder to the $60~\mbox{grams}$ of freshly washed particles. This results in enhanced defect formation within the lattice near region II even without frequency modulation of the container~\cite{sandpaper}; frequency modulation of these particles fully melts the pattern (see Fig.~\ref{comparison}(b)), creating a disordered liquid-like pattern in which lattice elements are no longer confined to oscillate near their equilibrium positions. We hypothesize that the friction controls the mode amplitude by damping the in-plane sliding motion of the peaks against the container bottom. 

In the experiments we cannot vary the friction in a controlled way, but this can be done in molecular dynamics (MD) simulations. We previously found that an MD code for inelastic hard spheres yielded square patterns like those in Fig.~\ref{largelattice}(a)~\cite{bizon98a}. With a friction coefficient $\mu=0.5$, the spatial patterns and their wavelengths obtained from MD simulations were in good accord with the experiments for a wide range of $\Gamma$ and $f_d$. We find that melting in the MD simulation occurs rapidly when the friction coefficient between the grains and the plate is reduced to zero, as Fig.~\ref{comparison}(c) illustrates. In this melting process the correlation length of the pattern $\xi$ decreases from about $5a$ to $2a$ (see inset of Fig.~\ref{decorr}). ($\xi$ was obtained by fitting an exponential to the envelope of the azimuthal average of the 2D autocorrelation function.) After reaching a minimum in about 30 plate oscillations, $\xi$ oscillates about $2a$ as defects continue to appear and disappear in the liquid-like pattern. Since the melting of the granular lattice is driven by the growth of the single excited mode, we cannot compare our observations with the well-known 2D melting theory of Halperin and Nelson~\cite{halAnel}, which requires uniform (thermal) heating of all modes.

{\em Lindemann melting criterion.} In simulations of 2D crystalline lattices, melting has been found to occur when the Lindemann ratio, $\gamma_M=\langle|u_{m}-u_{n}|\ ^{2}\rangle/a^{2}$, exceeds 0.1~\cite{bedanov,zheAear}; here $u$ denotes displacements of atoms from lattice sites, and the average is taken over all nearest neighbors $m$ and $n$.  The time evolution of $\gamma_M$ computed for the granular lattice for different values of $\mu$ is shown in Fig.~\ref{decorr}.  We find that the lattice melts ($\xi$ reaches its plateau region) when $\gamma_M$ increases through the value of $0.1$, just as in 2D crystals.  For $\mu=0.5$, the amplitude of the excited mode is too small for the lattice to melt, while for $\mu=0.1$, the pattern oscillates with increasing amplitude until local melting events disrupt the long-range order (Fig.~\ref{decorr})---such behavior is analogous to the dynamics seen in the experiment near the peak of region II. Thus, the Lindemann ratio can be used as a predictive criterion for the loss of order in the square patterns.

{\em Conclusions.} We have shown a close correspondence between the square patterns in a vibrated granular layer and a discrete set of harmonically coupled elements in a 2D lattice. The applicability of the Lindemann criterion in the nonequilibrium system further confirms the utility of the lattice dynamics framework. In addition, we conjecture that the stability exhibited by the nonequilibrium granular lattice pattern is an example of the ``generalized rigidity'' typically found in equilibrium lattice systems and discussed by Anderson~\cite{andAste}. Whether such concepts and predictive criteria can be applied to other nonequilibrium systems is an open question. For example, it would be interesting to see if other nonequilibrium patterns that display excited modes~\cite{othersystems} exhibit a dispersion relation and a melting scenario like that we have observed. Finally, we note that our observations of the effect of friction on melting of the lattice could help guide the development of kinetic and hydrodynamic theory of granular media, where the role of friction is not well understood. 

\begin{acknowledgments}
We thank Paul Umbanhowar for helpful suggestions.  This work was
supported by the Engineering Research Program of the Office of Basic
Energy Sciences of the U.S. Department of Energy under grant number DE-FG03-93ER14312.
\end{acknowledgments}

\bibliography{meltbib2}

\begin{thebibliography}{14}
\bibitem[{mod()}]{modenote}
\bibinfo{note}{The longitudinal $(1,\pm1)$ as well as the $(1,0)$ and $(0,1)$
  modes are absent, which indicates that the effective spring is stiff in those
  directions.}
\end{thebibliography}

\end{document}